# Oscillatory Hall effect from magnetoelectronic coupling in flexoelectronic silicon

Paul C. Lou[1], Ravindra G. Bhardwaj[1], Anand Katailiha[1], W.P. Beyermann[2], and Sandeep Kumar[1,3*]

[1] Department of Mechanical Engineering, University of California, Riverside, CA 92521, USA

[2] Department of Physics and Astronomy, University of California, Riverside, CA 92521, USA

[3] Materials Science and Engineering Program, University of California, Riverside, CA 92521, USA

[*] Corresponding author

Email: sandeep.suk191@gmail.com

**Abstract**

The magnetoelectronic coupling can be defined as cross-domain coupling between electronic and magnetic properties, where modulation in magnetic properties changes the electronic properties. In this letter, an explicit experimental evidence of magnetoelectronic coupling is presented, which is uncovered from oscillatory Hall effect response in Hall measurement. The strain gradient in a MgO (1.8 nm)/p-Si (~400 nm) freestanding sample leads to transfer of electrons (~$5\times10^{18}$ cm$^{-3}$) from valence to conduction band due to flexoelectronic charge separation in the p-Si layer. The resulting flexoelectronic polarization gives rise to temporal magnetic moment from dynamical multiferroicity. The external magnetic field changes the net temporal magnetic moment, which causes modulations in charge carrier concentration and oscillatory Hall effect. The period of oscillatory Hall response is 1.12 T, which is attributed to the magnitude of temporal magnetic moment. The discovery of oscillatory Hall effect adds a new member to the family of Hall effects.

## I. Introduction

In a recent discovery, a large phonon magnetic moment (1.2 μB/atom) was reported in Si thin film under an applied strain gradient. The large magnetic moment was attributed to the dynamical multiferroicity[1-4], which can be described as:

$$M_t \propto P_{Flexoelctronic} \times \partial_t P \quad (1)$$

where $M_t$, $P$ and $P_{Flexoelectronic}$ are temporal magnetic moment and polarization of optical phonons and the flexoelectronic polarization. The flexoelectronic polarization arises due to charge carrier transfer from metal layer to doped semiconductor layer in a metal/semiconductor heterostructure under an applied strain gradient; as demonstrated recently[5]. As a consequence, the flexoelectronic polarization is proportional to gradient of the charge carrier concentration ($n'$) and equation 1 can be rewritten as:

$$M_t \propto n' \times \partial_t P \quad (2)$$

This equation describes inhomogeneous magnetoelectronic multiferroicity and magnetoelectronic electromagnon (magneto-active phonon); as demonstrated recently[6].

The equation 2 also describes possibly the magnetoelectronic coupling in the materials. The magnetoelectronic coupling can be described as cross-correlation between the magnetic properties (temporal magnetic moment) and the electronic properties (charge carrier concentration), which is underlying cause of recently demonstrated magnetoelectronic electromagnon[6]. In materials having magnetoelectronic coupling, an external magnetic field applied at an angle to the direction of the temporal magnetic moment will give rise to precession of the temporal magnetic moment. As a consequence, the net temporal magnetic moment orthogonal to the polarization of optical phonons and flexoelectronic polarization will become smaller. Hence, the applied magnetic field will change the electronic charge separation (or flexoelectronic polarization) as well as the charge carrier concentration assuming polarization of phonons does not change as shown in Figure 1 (a) and can be described as:

$$(\boldsymbol{M}_t \pm \mathbf{B}) \propto (\boldsymbol{n} \pm \Delta \boldsymbol{n})' \times \partial_t \boldsymbol{P} \qquad (3)$$

where **B** and $\Delta \boldsymbol{n}$ are external magnetic field and change in charge carrier concentration, respectively. Since, the external magnetic field changes the electronic properties of the material, we call it magnetoelectronic coupling. This is analogues to magnetoelectric coupling, which allow electric control magnetic behavior in multiferroic materials[7,8]. In this letter, we present an experimental evidence of the oscillatory Hall effect response due to magnetoelectronic coupling. The magnetoelectronic coupling lead to change in charge carrier concentration as a function of magnetic field.

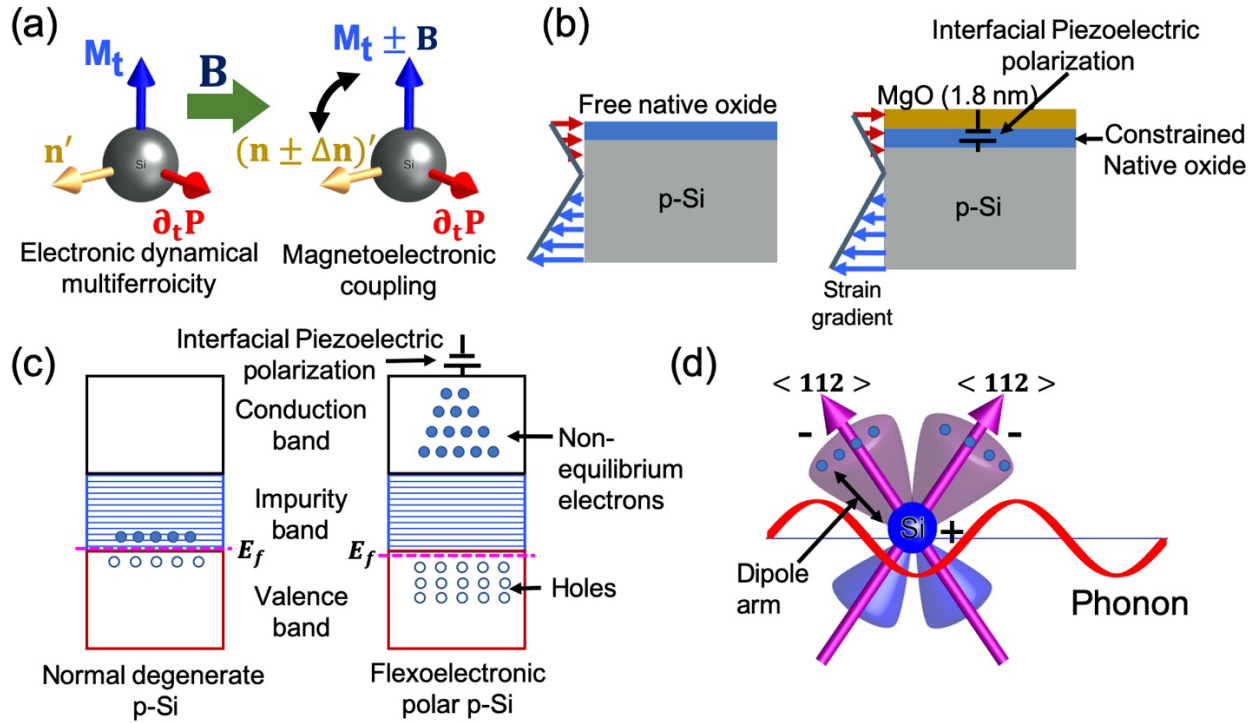

Figure 1. (a) Schematic showing the magnetoelectronic coupling from an externally applied magnetic field due to electronic dynamical multiferroicity. (b) Schematic showing the interfacial piezoelectric polarization in native oxide under an applied strain gradient, which acts on the degenerately doped Si layer. (c) Schematic showing the charge carrier distribution due to interfacial piezoelectric like effect leading to non-equilibrium electrons in the conduction band giving rise to metastable flexoelectronic polar p-Si layer. (d) Schematic showing the deformed charge distribution along the close packed directions <112> in [110] plane, which give rise to the electronic polarization and temporal magnetic moment of magnetoelectronic electromagnon due to superposition from phonons.

## II. Experimental results

The flexoelectronic charge carrier transfer has so far been demonstrated primarily in the metal/doped semiconductor heterostructures under applied strain gradient.

However, the response from the metal layer is superimposed on the response from the semiconductor layer, which makes it difficult to segregate the responses. The potential flexoelectronic effect has also been demonstrated in MgO/p-Si (degenerate) bilayer structures but the mechanism was not identified since there is no charge carrier metal source in the new bilayer structures[9]. This led us to choose a similar bilayer configuration for this study since the transport response will only arise from the degenerately doped Si layer.

The charge carrier concentration is usually measured using Hall effect. Hence, the magnetoelectronic coupling and change in charge carrier concentration can also be uncovered using Hall resistance measurement. For the experimental setup, we take a degenerately doped (Boron) 2 µm thick p-Si (0.001-0.005 $\Omega$-cm) silicon on insulator (SOI) wafer (commercially available). We, then, reduce the thickness of the device p-Si layer to ~ 400 nm using successive oxidation and hydrofluoric acid (HF) etching[10]. The device p-Si layer thickness is reduced to achieve the larger strain gradient. The device layer is, then, patterned and etched in a Hall bar configuration using photolithography and Si deep reactive ion etching, respectively[9]. The sample area is made freestanding using HF vapor etch of the buried oxide layer underneath the device layer. We take two samples from the same part of the wafer in order to have similar initial doping characteristics. The sample 1 is the control sample where no further processing is carried out. For the sample 2, we deposited 1.8 nm of MgO on top of the p-Si layer[9]. The thickness of native oxide layer is expected to be 3.7 nm[11].

The freestanding p-Si layer will have residual stress and buckling will induce strain gradient in the bulk of the sample. In the sample 1, the strain gradient may lead to small charge carrier separation due to gradient in the band structure. But, the native oxide layer is free to deform and will not influence the overall response as shown in Figure 1 (b). Whereas, the native oxide in sample 2 is constrained by MgO top layer and is not allowed to freely expand or contract. As a result, the strain gradient will lead to piezoelectric like response[12] from the native oxide layer on top of p-Si as shown in Figure 1 (b). The piezoelectric like response from the native oxide layer arises, potentially, due to non-stoichiometry and dangling bonds. It is noted that deposition of any layer on top of Si with native oxide will constrain the native oxide layer and MgO layer is not expected to contribute towards the interfacial response[13]. However, the deposition of the MgO layer is undertaken using RF sputtering, which may lead to charge accumulation at the surface. It is noted that strain gradient is not expected to induce any phase transition in the Si[11,14]. In a degenerately doped p-Si, the impurity states give rise to a continuous impurity band and Fermi level will be near the edge of valence band as shown in Figure 1 (c). Now, the interfacial piezoelectric like response can be considered as a gate bias on the Si layer. The free charge carrier distribution inside the Si layer will deform to neutralize the interfacial piezoelectric like response and surface charge accumulation in the second case as shown in Figure 1 (c). The charge carriers are expected to jump to conduction band as shown in Figure 1 (c) leaving behind holes in the valence band. This increase in free charge carriers will increase the conductivity as well as Hall resistance. This proposed behavior has been reported in a MgO/n-Si (2 µm) sample where current dependent response showed an increase in charge carrier concentration from ~-5.9×10$^{19}$ cm$^{-3}$ at 1 mA of current to ~-8×10$^{19}$ cm$^{-3}$ at 5 mA due to increased buckling[1]. Similarly, the resistivity of the sample also reduced from ~2.44×10$^{-5}$ Ω-m (34.95 Ω) at 1 mA to

~1.5×10$^{-5}$ Ω-m (24.73 Ω) at 5 mA[1]. The non-equilibrium charge carrier transfer from the interior of the atom to outer edge of the atom will lead to deformed charge distribution around the atom as shown in Figure 1 (d) and as stated earlier. The resulting partial ionization of the p-Si layer from deformed charge distribution will give rise to a metastable state having dipole like behavior as shown in Figure 1 (d). We call it flexoelectronic polarization because it is an electronic response to a strain gradient. The flexoelectronic polarization will be equal and opposite to the interfacial piezoelectric like effect. Then, the superposition of the flexoelectronic polarization and circularly polarized phonons will give rise to dynamical multiferroicity; essential for magnetoelectronic behavior. The nonequilibrium free charge carrier will preferentially reside along the close-packed directions in the cross-section of the sample such as <112> in the [110] plane giving rise to temporal magnetic moment along <111> directions as reported earlier[1,6] and as shown in Figure 1 (d).

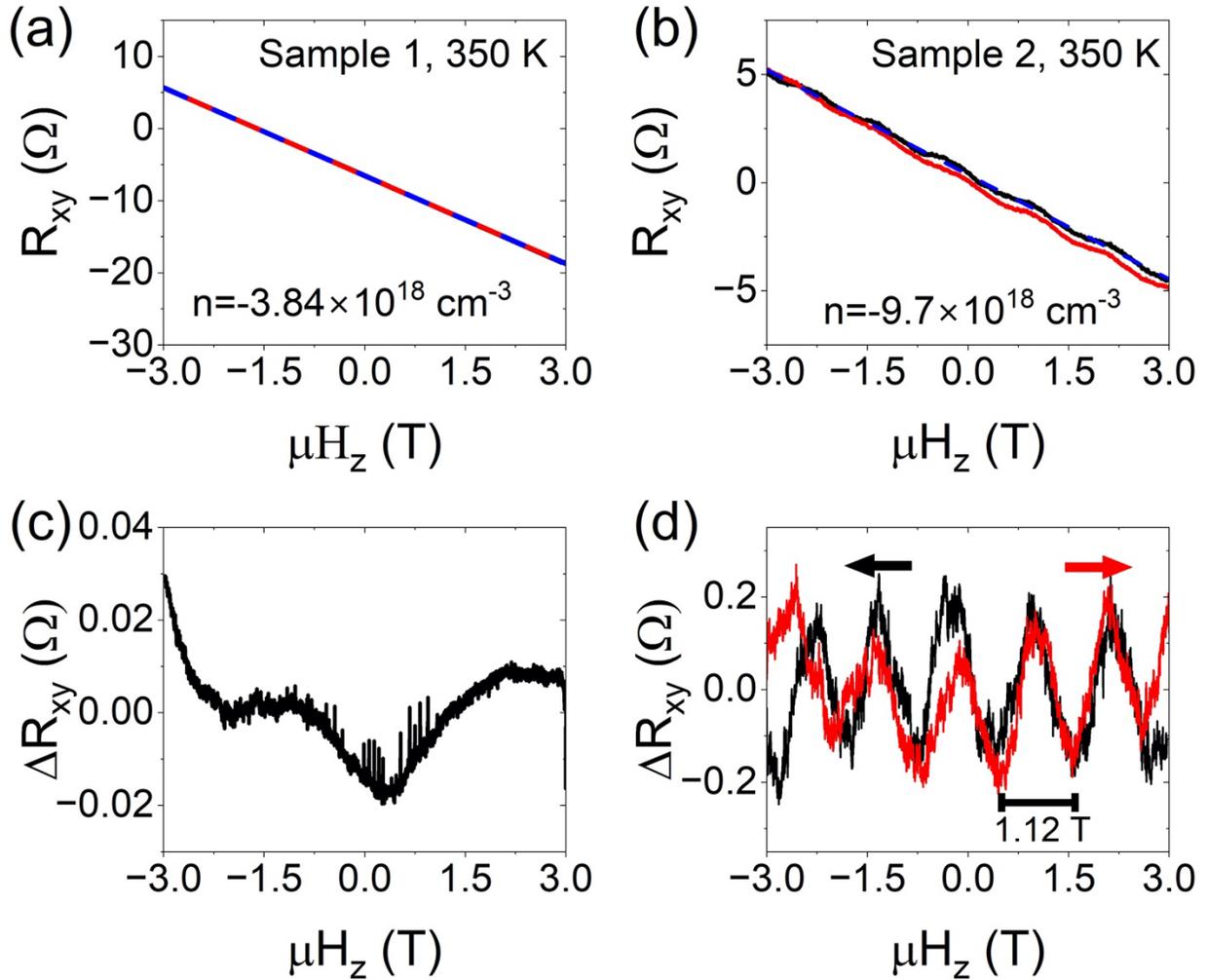

Figure 2. (a) The Hall response measured in control sample 1 at 350 K. (b) The Hall response in sample 2 at 350 K. (c) The residual response from random noise in control sample 1 at 350 K. (d) And the oscillatory component of the Hall response in sample 2 showing a period of 1.12 T.

In the first set of experiments, we measured the Hall response as a function of magnetic field from 3 T to -3 T at 350 K in control sample 1 (200 µA) and sample 2 (500 µA) as shown in Figure 2 (a,b). An alternating current (a.c.) bias is applied using Keithley 6221 current source and response is measured using Stanford Research Systems SR830 lock-in amplifier. The different biasing current are used to account for the different resistivities of the samples. At 350 K, the Hall measurement in sample 1 shows a negative Hall resistance and charge carrier concentration of $-3.84 \times 10^{18}$ cm$^{-3}$ as shown in Figure 2 (a). The negative sign of the Hall resistance is surprising at the measured charge carrier concentration and can be attributed to Fermi level being near the valence band edge as well as from strain gradient induced gradient in band structure. Zhang and Chang[15] proposed that charge migration may enhance the flexoelectric response in Si. The observed sign of Hall resistance in control sample 1 may also arise from charge migration behavior. The residual response from the line fit is shown in Figure 2 (c), which exhibits noise floor of the measurement. As compared to control sample 1, the sample 2 exhibits an anomalous oscillatory Hall response as shown in Figure 2 (b). Using a line fit, we extract the linear Hall resistance and the oscillatory response as shown in Figure 2 (b,d). The Hall resistance is negative and the charge carrier concentration is estimated to be $-9.7 \times 10^{18}$ cm$^{-3}$. The difference in the charge carrier concentration between sample 1 and sample 2 is $4.86 \times 10^{18}$ cm$^{-3}$. This difference is attributed to the transfer of charge carriers from the interior of the atom to the conduction band in reaction to interfacial piezoelectric like response from constrained native oxide; as hypothesized and as shown in Figure 1 (c). This shows that in spite of charge neutrality an electronic polarization (flexoelectronic) can be achieved in an inhomogeneous system.

The oscillatory response seems to be triangular as shown in Figure 2 (d). A negative slope will add and positive slope will act opposite to the to the negative Hall response. We estimate the charge carrier concentrations to be $-15.6 \times 10^{18}$ cm$^{-3}$ and $-7.0 \times 10^{18}$ cm$^{-3}$ for positive and negative slopes. This behavior indicates fluctuation of free charge carrier concentration due to an applied magnetic field, which we call magnetoelectronic coupling; as hypothesized earlier. The superposition of flexoelectronic polarization from non-equilibrium free charge carrier and circularly polarized phonons give rise to temporal magnetic moment of resulting magnetoelectronic electromagnon. An externally applied magnetic field act on the temporal magnetic moment and, as a consequence, the charge carrier concentration is also modified; as hypothesized earlier. From the oscillatory response, the period of the oscillation is estimated (using sine fit) to be ~1.12 T as shown in Figure 2 (d). We, therefore, propose that the period of oscillation (1.12 T) is the magnitude of the temporal magnetic moment of magnetoelectronic electromagnon, which is larger than the value (0.7 T) reported in case of 2 μm thick n-Si sample[1]. The oscillatory component of the response is ~4% of the transverse resistance value at 3 T (~5 Ω). One can argue that the response is noise from measurement. However, a random noise is not expected to be a function of magnetic field with period of ~1.12 T. In addition, the amplitude of the oscillatory response (0.2 Ω) is constant for complete range of magnetic field as shown in Figure 2 (d). Further, the oscillatory response (~4%) is an order of magnitude larger than the noise (<0.1%) measured in the sample 1 as shown in Figure 2 (c) whereas the order of the Hall response is similar in sample 1 and 2 as shown in Figure 2 (a,b). Hence, the observed oscillatory response is not expected to arise from random noise. We also measured the longitudinal response between 1 T and -1 T at 350 K. However, the magnetoresistance is too small to uncover any effect of oscillatory response on longitudinal resistance.

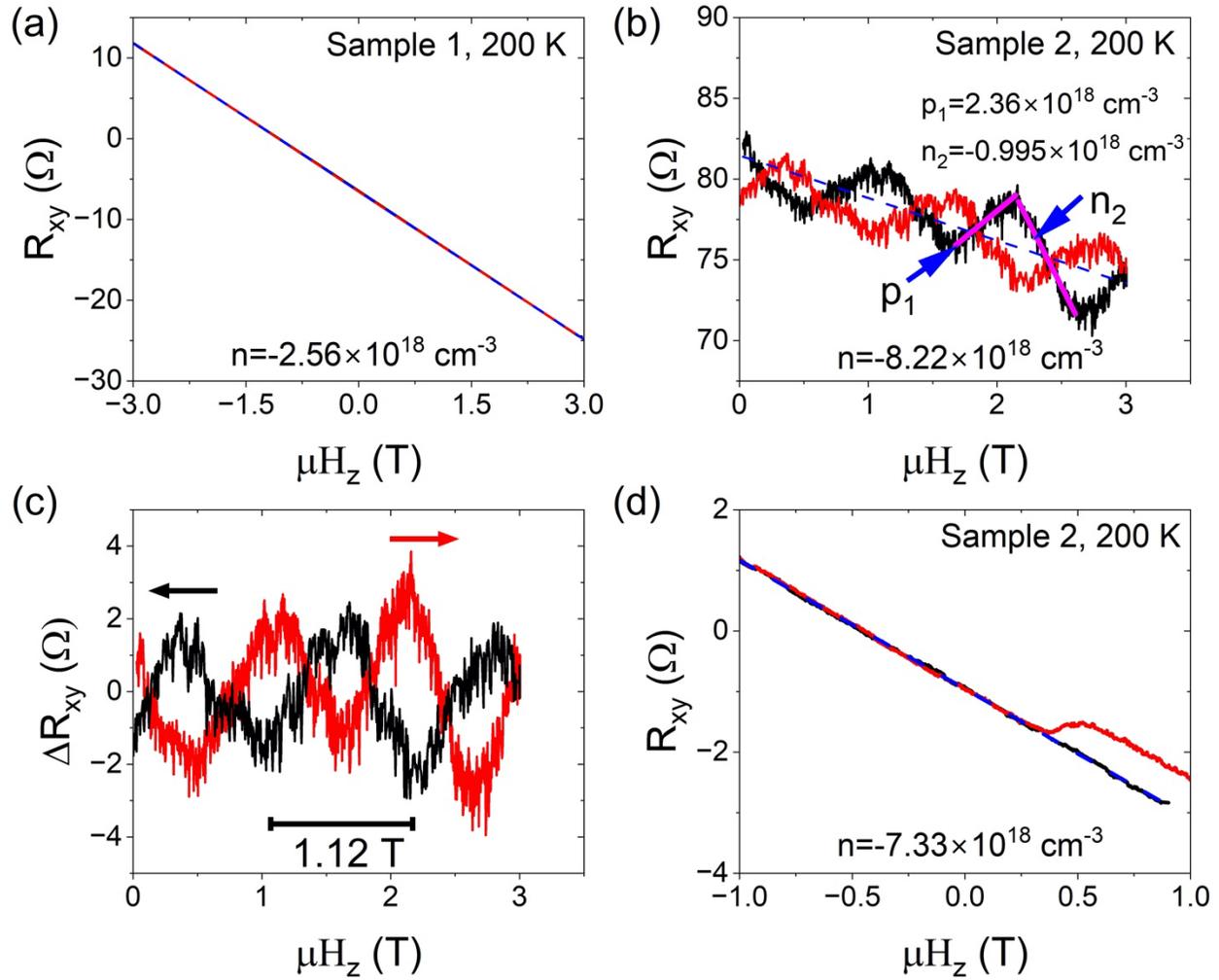

Figure 3. (a) The Hall response measured in control sample 1 at 200 K. (b) The Hall response in sample 2 at 200 K from 0 T to 3 T. (c) The oscillatory component of the Hall response in sample 2 showing a period of 1.12 T. (d) The second Hall response in sample 2 at 200 K from 1 T to -1 T.

We, then, measured the Hall response in control sample 1 at 200 K and observed a negative Hall resistance as shown in Figure 3 (a). The charge carrier concentration in the control sample 1 is estimated to be $-2.56\times 10^{18}$ cm$^{-3}$ as shown in Figure 3 (a). We also measured the Hall response in sample 2 from 0 T to 3 T at 200 K, where largest field is larger than the expected period of oscillation. At 200 K, the oscillatory Hall response was pronounced as shown in Figure 3 (b). The linear Hall resistance is negative and the charge carrier concentration is estimated to be -8.22×10$^{18}$ cm$^{-3}$. This is larger than the charge carrier concentration in sample 1 by 5.66×10$^{18}$ cm$^{-3}$., which is a clear evidence of flexoelectronic polarization in Si due to interfacial piezoelectric like effect as hypothesized.

Similar to 350 K, the oscillatory response at 200 K has a period of ~1.12 T as shown in Figure 3 (c). However, the oscillatory response has an opposite sign based on direction of field sweep as shown in Figure 3 (c). From the Hall response in Figure 3 (b), we observe that the Hall resistance changes sign from positive to negative. A linear fit into two opposite slopes show the charge carrier concentration fluctuates due to applied magnetic field as shown in Figure 3 (b). The estimated charge carrier concentrations are -0.99×10$^{18}$ cm$^{-3}$ and +2.36×10$^{18}$ cm$^{-3}$ for negative and positive slopes as shown in Figure 3 (b). This change shows that a large number of charge carriers move between conduction band and valence band during oscillatory response. This behavior supports our primary hypothesis of flexoelectronic charge separation depicted in Figure 1 (c) as well as magnetoelectronic coupling since magnetic field lead to change in the charge carrier concentration. The oscillatory component is absent completely in control sample 1 at both 350 K and 200 K, which proves our hypothesis.

We hypothesized that if the externally applied magnetic field is smaller than the period of oscillation then the oscillatory behavior will disappear. Hence, we measured the Hall response in sample 2 again for an applied magnetic field from 1 T to -1 T at 200 K as shown in Figure 3 (d). We do not observe any oscillatory response in this measurement, which supports our hypothesis. However, we observe a hysteretic response (possibly anomalous Hall response) at 200 K as shown in Figure 3 (b), which can also potentially arise from magnetoelectronic coupling and temporal magnetic moment of magnetoelectronic electromagnon. In this measurement, the charge carrier concentration is estimated to be -7.34×10$^{18}$ cm$^{-3}$. It is smaller than the charge carrier concentration estimated using previous Hall response in Figure 3 (b) by 0.89×10$^{18}$ cm$^{-3}$. This difference showed that larger magnetic field potentially increases the charge carrier concentration significantly, which is clear proof of hypothesized magnetoelectronic coupling.

It is noted that the transverse resistance in the first measurement on sample 2 is higher by an order of magnitude and the underlying reason of this difference is currently unknown. Further, in the first Hall response measurement, we started the magnetic field at 0 T and then sweeping it to 3 T rather than starting at 3 T. In spite of that, the oscillatory response is observed in this measurement. It is noted that measurement at 350 K is followed by this measurement at 200 K without breaking the sample chamber vacuum. As a consequence, the residual magnetic moment of magnetoelectronic electromagnon due to high magnetic field at 350 K may induce oscillation at 200 K even though we do not start the measurement at high magnetic field. The second Hall measurement in sample 2 at 200 K was carried out separately after the sample was out from the chamber for long duration.

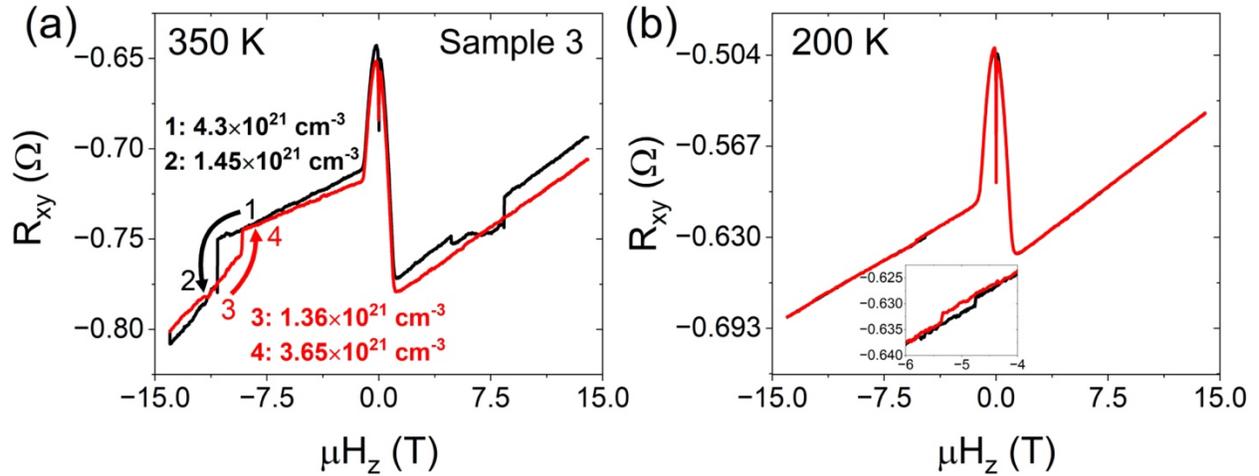

Figure 4. The Hall response in the sample 3 from 14 T to -14 T at (a) 350 K and (b) 200 K.

The flexoelectronic effect is previously reported in metal/doped semiconductor heterostructures. The measurement in sample 1 and sample 2 demonstrated dynamical multiferroicity and magnetoelectronic coupling in Si. To further support our argument, we take a Py (25 nm)/MgO (1.8 nm)/p-Si (400 nm) sample (sample 3). We measured the Hall response at 350 K as a function of magnetic field from 14 T to -14 T and at an applied current bias of 2 mA shown in Figure 4 (a). The measured Hall response is positive. It suggests that the dominant charge carriers are holes, which is unexpected since electrons are the dominant charge carrier in Py. Based on the previous resistance measurements, the Py resistivity at 350 K is expected to be $5.72\times10^{-7}$ $\Omega$m[5] and p-Si resistivity is expected to be $1.36\times10^{-4}$ $\Omega$m (from sample 1). However, the resistance of the heterostructure sample is found to be 32.3 $\Omega$. As a consequence, the p-Si resistivity in the heterostructure need to be $8.94\times10^{-6}$ $\Omega$m, assuming Py resistivity remains same. The decrease in resistivity of p-Si layer is attributed to the flexoelectronic charge carrier transfer from Py to p-Si layer, which leads to electron deficiency in the Py layer and is also the underlying cause of the positive Hall resistance[5]. In the Hall measurement, we observe a hysteretic response as shown in Figure 4 (a). Based on the measured Hall resistances, the charge carrier concentration reduces from $4.3 \times 10^{21}$ cm$^{-3}$ to $1.45 \times 10^{21}$ cm$^{-3}$ at the magnetic field -10.8 T as shown in Figure 4 (a). In the inverse magnetic sweep, the charge carrier concentration increases from $1.36 \times 10^{21}$ cm$^{-3}$ to $3.65 \times 10^{21}$ cm$^{-3}$ at the magnetic field 9.2 T as shown in Figure 4 (a). The large magnetic field leads to the spin of charge carrier aligning with the external magnetic field. As a consequence, the reduction in the charge carrier concentration is attributed to the transfer of electrons below the Fermi level and in turn reducing the hole population. This behavior is similar to the oscillatory behavior observed in sample 2 except there are no oscillations in the response. The absence of oscillatory response is attributed to the Py dominated Hall response, which masks the response from the Si layer. The additional effects due to interlayer coupling and proximity effect may also diminish the oscillatory response. The Hall resistance at 200 K is also positive due to flexoelectronic charge transfer as shown in Figure 4 (b). The hysteresis behavior at 200 K is at lower magnetic field and weaker in magnitude as compared to 350 K as shown in Figure 4 (b) inset. This is attributed to freezing of the magnetoelectronic electromagnon and resulting decrease in the temporal magnetic moment. This behavior is opposite of sample 2 where oscillatory response is larger at 200 K. The flexoelectronic effect in both sample 2 and 3 is due to different mechanisms. The Si layer is charge neutral in sample 2 whereas it is not in sample 3 since there are excess charge carrier from the Py layer in sample 3. As a consequence, the different behavior emerges from the lack of charge neutrality in sample 3. However, studies are needed to elucidate it further.

## III. Conclusion

In conclusion, we presented the experimental evidence of magnetoelectronic coupling in flexoelectronic Si. The magnetoelectronic coupling arises due to superposition of flexoelectronic polarization and circularly polarized phonons that also give rise to temporal magnetic moment of magnetoelectronic electromagnon. The externally applied magnetic field modulates the temporal magnetic moment, which in turn changes the charge carrier concentration. This magnetoelectronic coupling leads to an oscillatory Hall effect response in the Hall measurement. The oscillatory response in the Hall resistance arises due to back and forth transfer of charge carrier from interior of the atom to the conduction band due to modulation of temporal magnetic moment from magnetic field. The period of the oscillatory Hall response is ~1.12 T, which is expected to corresponds to the magnitude of temporal magnetic moment of magnetoelectronic electromagnon. Though we presented the evidence of magnetoelectronic coupling but a complete quantitative description and microscopic origin of the behavior will need further experimental and theoretical studies. Further, the discovery of oscillatory Hall effect adds a new member to the family of Hall effects from different origin.

**Acknowledgement**

The fabrication of experimental devices was completed at the Center for Nanoscale Science and Engineering at UC Riverside. SK acknowledges a research gift from Dr. Sandeep Kumar.